# Impact of Valence States on Superconductivity of Oxygen Incorporated Iron Telluride and Iron Selenide Films

D. Telesca[1], Y. Nie[1, 2], J. I. Budnick[1, 2], B. O. Wells[1, 2], and B. Sinkovic[1]
[1]*Department of Physics, University of Connecticut, Storrs, CT 06269, USA and*
[2]*Institute of Materials Science, University of Connecticut, Storrs, CT 06269-3136, USA*

We report on the local electronic structure of oxygen incorporated FeTe and FeSe films and how this relates to superconductivity observed in these films. In the case of FeTe, intially grown films are measured to be non-superconducting, but become superconducting following oxygen incorporation. In FeSe the opposite happens, initially grown films are measured to be superconducting, but experience a quenching of superconductivity following oxygen incorporation. Total Fluorescence Yield (TFY) X-ray absorption experiments show that oxygen incorporation changes the initial Fe valence state in both the initially grown FeTe and FeSe films to mainly $Fe^{3+}$ in the oxygen incorporated films. In contrast we observe that while Te moves to a mixed $Te^0/Te^{4+}$ valence state, the Se always remains $Se^0$. This work highlights how different responses of the electronic structure by the respective chalcogenides to oxidation could be related to the mechanisms which are inducing superconductivity in FeTe and quenching superconductivity in FeSe.

The discovery of superconductivity in the iron pnictide $LaFeAsO_{1-x}Fe_x$[1] (denoted as 1111 compound) has led to intense investigation of other Fe based superconductors. Hsu et al. discovered superconductivity in FeSe (Tc = 8 K[2]) which has a simple PbO-type structure and a similarity to the critical $FeAs_4$-tetrahedra layers found in all iron-based superconductors. The critical temperature was increased through the partial substitution of Te for Se to a maximum value of $T_c$ = 14 K in $FeSe_xTe_y$ (x=y=0.5)[3] and through the application of high pressure, achieving Tc = 37 K.[4] It has been found that superconductivity disappears at y = 1 in FeTe.[5,6] In addition, it was demonstrated that oxygen poisoning of $Fe_{1.01}Se$ results in a less sharp transition when contrasted to the sharp transition as compared to oxygen free $Fe_{1.01}Se$ near.[7]

Due to the isostructural and isoelectronic nature of FeSe and FeTe, it has been surprising that no superconductivity has been observed in single crystal FeTe. Y. Mizuguchi et al.[8] have demonstrated that post growth oxygen annealing is an effective method to induce superconductivity in bulk polycrystalline $FeTe_{1-x}S_x$, but attempts to induce superconductivity in oxidized bulk polycrystalline FeTe were unsuccessful. We recently reported superconductivity in the FeTe film system by the incorporation of oxygen through post growth oxygen annealing.[9] Other reports of FeTe films exhibiting superconductivity also exist in the literature.[10,11]

The question of what parameters control the appearance of superconductivity in the iron superconductors is under intense study. Key ingredients considered include the doping level, the detailed crystal structure, and the relationship to long range ordered magnetism. It has been demonstrated that the tetragonal to orthorhombic structural transition and the long-range antiferromagnetic transition must both be suppressed before the optimum $T_c$ is obtained in any of the FeAs-based systems.[12] This suggests that superconductivity and long-range antiferromagnetic ordering strongly interact, and in fact compete with each other.[13] Studies furthering the understanding of this issue were performed on the $Ba(Fe_{1-x}M_x)_2As_2$ (M = Co, Ni, Rh) systems which have phase diagrams[14] that show regions where superconductivity and long-range antiferromagnetic order appear to coexist.[13] A relationship between superconductivity and magnetism has also been demonstrated in the Fe-chalcogenide superconductors.[15] The effect of oxygen incorporation into non-superconducting FeTe and superconducting FeSe adds important information on the underlying phase diagram for superconductivity in this family of compounds.

In this Rapid Communication we report changes to the local electronic structure of FeSe and FeTe films resulting from oxygen incorporation. These changes are examined in light of the corresponding resistivity measurements which show a suppression of the superconducting transition in oxygen incorporated FeSe films and the onset of superconductivity in oxygen incorporated FeTe films. X-ray absorption (XAS) measurements indicate that the Fe of both parent films experiences a similar nominal valence change whereas the valence states of the respective chalcogenides (Te,Se) have different responses.

FeTe and FeSe films, 80 ± 15 nm thick, were grown and characterized in the same manner as described in our previous work.[9] XAS experiments were performed at the National Synchrotron Light source at the U4B beam line. All the XAS data were taken with normal photon beam incidence, with a beam spot of 1x3 $mm^2$ and photon energy resolution of 0.34 eV for the Fe-L edges and 0.42 for the Te-M edges. All XAS data were normalized by the photon flux recorded with a gold mesh. The XAS spectra were recorded simultaneously in total florescence yield (TFY) and total electron yield (TEY) modes, for probing the bulk and the surface of the films, respectively.

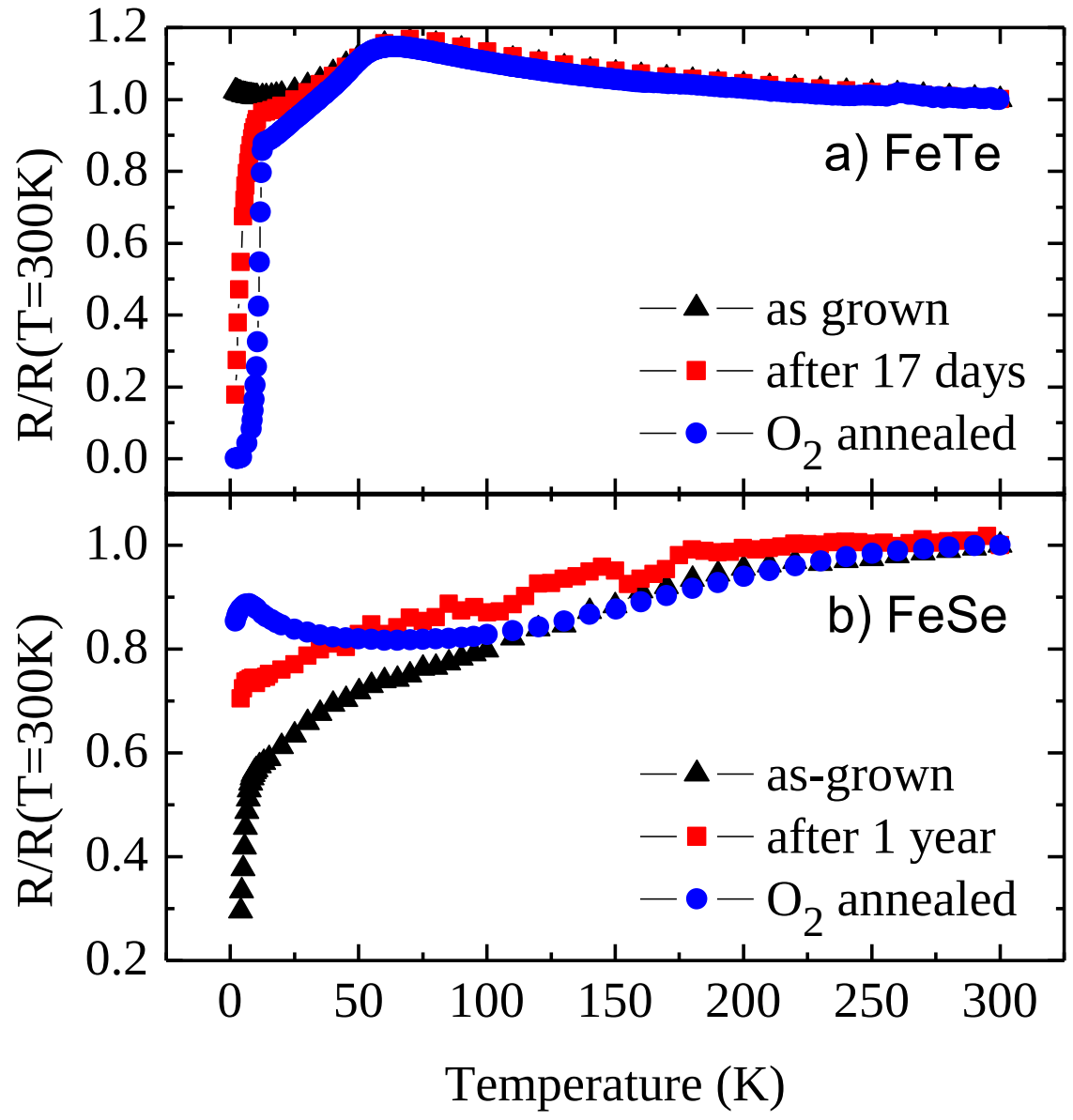

Figure 1: Resistivity measurements normalized to the value at 300 K for a) FeTe films with various amounts and types of oxygen exposure and b) FeSe films with various amounts and types of oxygen exposure

We first describe the transport state of the two Fe-chalcogenide films associated with different oxidation treatments. Figures 1a) and b) show resistivity measurements of the FeTe and FeSe films, respectively. As was observed in our previous work[9] and shown in Figure 1a) by the solid triangles, initially grown FeTe films exhibit a weekly metallic behavior. However, when the films incorporate enough oxygen, they become superconducting. Oxygen incorporation is accomplished either through a sufficient length of ambient air exposure[9] (solid squares) or low temperature oxygen annealing (solid circles), which consists of exposing the films to 100 mTorr of $O_2$, for 30 min, at 100°C. The opposite trend in transport behavior is observed in the isostructural FeSe film system. In Figure 1b) the solid black triangles show that initially grown FeSe films are superconducting, with a Tc ~ 8 K, consistent with PLD films grown in other work[16]. Oxygen incorporated FeSe films, experience a suppression of the superconducting state, either through long term exposure to air or $O_2$ annealing.

Fe-$L_{2,3}$ edge XAS measurements were taken of both non-superconducting and superconducting FeTe and FeSe films in TFY mode to determine the valence state of Fe associated with the different transport states of the respective films. Figure 2a) shows the Fe-L edges for an FeTe film that was exposed to air for 4 hours and measured to be non-superconducting; for an FeTe film that was exposed to air for 4 days and also measured to be non-superconducting; for an FeTe film exposed to air for 24 days and measured to be superconducting; and for an FeTe film $O_2$ annealed and measured to be superconducting. From the 4-hour-air-exposed spectrum, the most correct statement that can be made about the valence state for this stage of oxygen incorporation is that it is dominated by a mixture of $Fe^0$ and $Fe^{2+}$ (with spectral maximum located at 706.4 eV), along with some contribution of $Fe^{3+}$, which can be seen as the shoulder located at 708.7 eV (indicated by the arrow in Figure 2a). Separating the contributions of $Fe^0$ and $Fe^{2+}$ in such a mixed valence state is complicated by the overlap of energy positions of the maxima of $Fe^0$ and $Fe^{2+}$ at both the Fe-$L_3$ and –$L_2$ edges as can be demonstrated from experimental spectra of Fe and FeO (see for example T.J. Regan, et al.[17]). However, the contribution from $Fe^{2+}$ can still be detected by the shape of the $L_2$ edge, having a more extended multiplet structure (as for example in FeO) when compared to $Fe^0$. While proper identification of the initially grown valence state of Fe in FeTe films is important, it is the change of Fe to mostly 3+, discussed below, in the superconducting FeTe films that we focus on in this work.

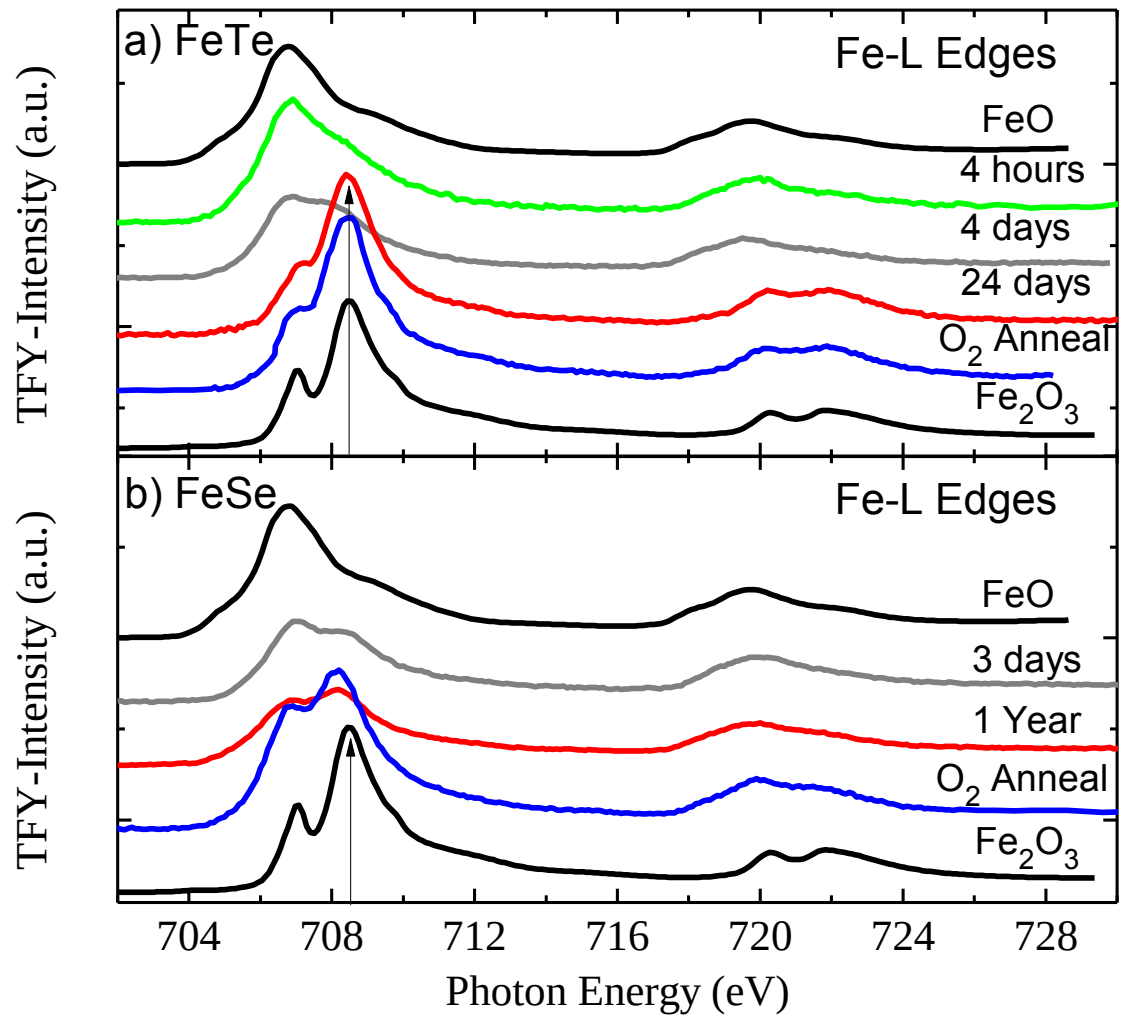

Figure 2: Fe L edge XAS-TFY spectra for a) FeTe films with various amounts and types of oxygen exposure and b) FeSe films with various amounts and types of oxygen exposure

The increase in intensity of the peak at 708.7 eV in the spectrum from the FeTe film exposed to air for 4-days we interpret as an indication of the increase of $Fe^{3+}$ states with continued exposure to air. A significant exposure to oxygen completes this trend and transforms the Fe XAS spectra of oxygen incorporated FeTe films (both 24-days-air-exposure and $O_2$ annealing) to the spectrum that closely resemble that of bulk $Fe_2O_3$, also shown in Figure 2a) for comparison. This indicates that the majority valence state of the iron in superconducting FeTe films is dominated by $Fe^{3+}$.

The changes in the Fe valence in FeSe films with oxygen treatment is also determined via XAS measurements. Figure 2b) shows the Fe-L edges for a 3-days-air exposed FeSe film that was measured to be superconducting; a 1-year-air exposed FeSe film that was measured to be non-superconducting; and an $O_2$ annealed (100 mTorr $O_2$, for 30 min. at 100°C) FeSe film measured to be non-superconducting.

Similar to the non-superconducting 4-day-air-exposed FeTe film, the spectrum from the superconducting 3-day-air-exposed FeSe film indicates that the Fe valence is most likely a mixture of $Fe^0$ and $Fe^{2+}$, along with some $Fe^{3+}$. The FeSe film was then observed to become non-superconducting by the same oxygen incorporation methods demonstrated to induce superconductivity in FeTe films (sufficient air exposure or $O_2$ annealing). The resulting Fe-$L_3$ spectra show the peak related to $Fe^{3+}$ at 708.7 eV (indicated by the arrow) increasing in intensity as compared to the 3-day-exposed FeSe film. However, the two main peaks at the Fe-$L_3$ edge do not show an intensity ratio that can be associated purely with the bulk $Fe_2O_3$ spectrum. Furthermore, the Fe-$L_2$ edge is more similar to the Fe-$L_2$ edge from bulk FeO, indicating a significant presence of $Fe^{2+}$. Therefore, an oxidized FeSe film does not have iron in a mostly $Fe^{3+}$ state. We conclude that while the same general trend in Fe valence change is observed in both FeTe and FeSe films, the rate at which these occur are different.

We now turn to our study of the electronic state of the chalcogens. Studies of the local electronic structure of Te in the FeTe films were performed to determine the impact of incorporated oxygen on the Te valence. Figure 3 shows the background subtracted Te-$M_{4,5}$ XAS-TEY and –TFY spectra for a set of FeTe films. Limited published Electron Energy Loss Spectroscopy (EELS) data indicate that $Te^0$ has an $M_5$ edge at ~572 eV and $M_4$ edge at ~583 eV[18,19] and that the $Te^{4+}$ $M_5$ and $M_4$ edges appear at respectively higher photon energies than their corresponding metallic edges[20]. The XAS-TEY spectra from 4-hour-exposed FeTe film shows peaks at 576.1 eV and 586.7 eV with smaller features on the higher energy side of each of these peaks. This indicates that the Te is primarily metallic. The XAS-TEY spectra of both the 24-days-air exposed and $O_2$ annealed superconducting films shows a dramatic increase of the higher energy peaks at 580.2 eV and 590.2 eV. This indicates that the Te in the surface of the superconducting FeTe films is mixed $Te^0/Te^{4+}$ valence. The corresponding TFY spectra in Figure 3 of the oxygen incorporated superconducting FeTe films confirms that this dramatic valence change to mixed $Te^0/Te^{4+}$ occurs through the majority of the films and not only at the surface.

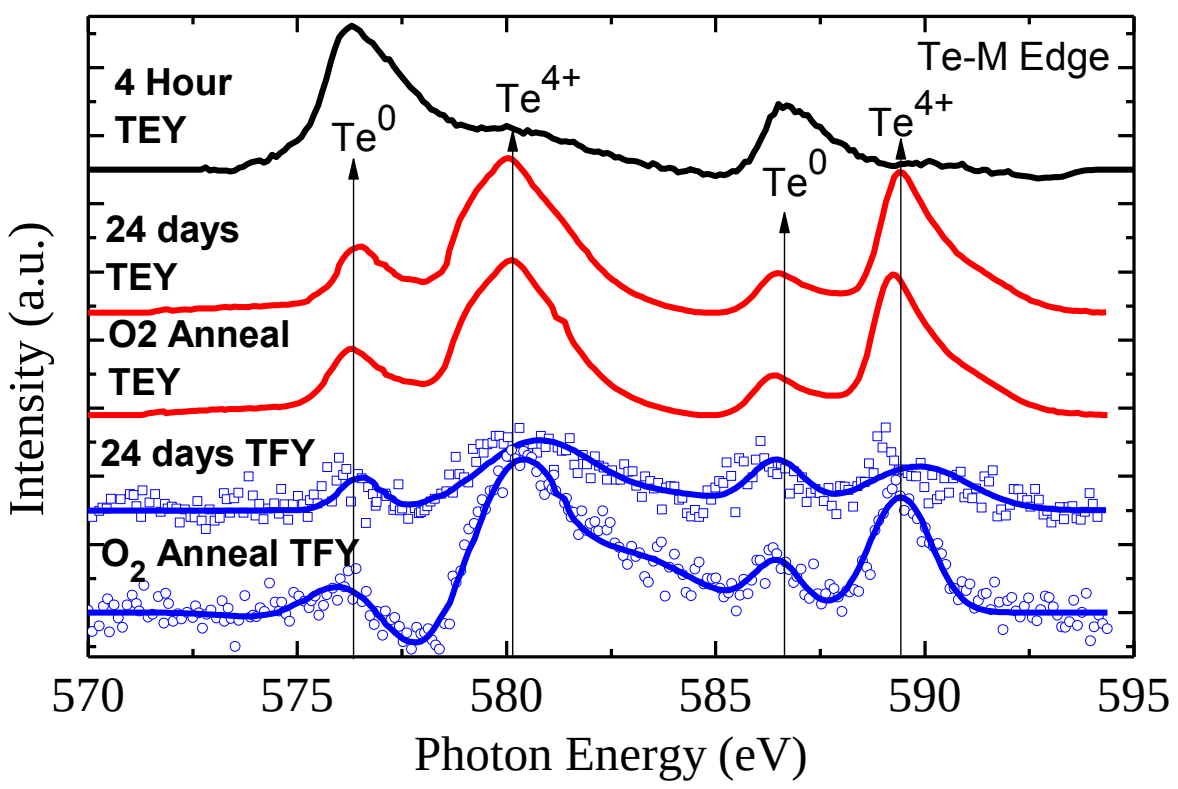

Figure 3: XAS-TEY Te-M edge spectrum from a non-superconducting 4-hour air exposed FeTe film and XAS-TEY and XAS-TFY Te-M edge XAS spectra from superconducting 24-days-air exposed and oxygen annealed FeTe films. The specta in the figure are background subtracted spectra. Open squares are the raw TFY data and the solid blue line is a guide to the eye for the raw data.

The study of the valence state of Se was performed with core-level XPS because the Se XAS edges were outside of the photon energy range available at the U4B beam line. Although the XPS is more surface sensitive technique than XAS-TFY, XPS spectra taken from in-situ grown FeSe films provide clear evidence what effect does oxygen incorporation has on the valence of Se.

Figure 4a) shows the Se-3d spectrum from a PLD grown Se reference film, an in-vacuo transferred FeSe film, and the spectrum following oxygen annealing of this film. The binding energies of the spectra were first calibrated from the Fermi Energy of the in-vacuo transferred films, for both the Se and FeSe films. For the $O_2$ annealed FeSe film, the binding energy was calibrated by the binding energy position of adventitious carbon at 284.6 eV.

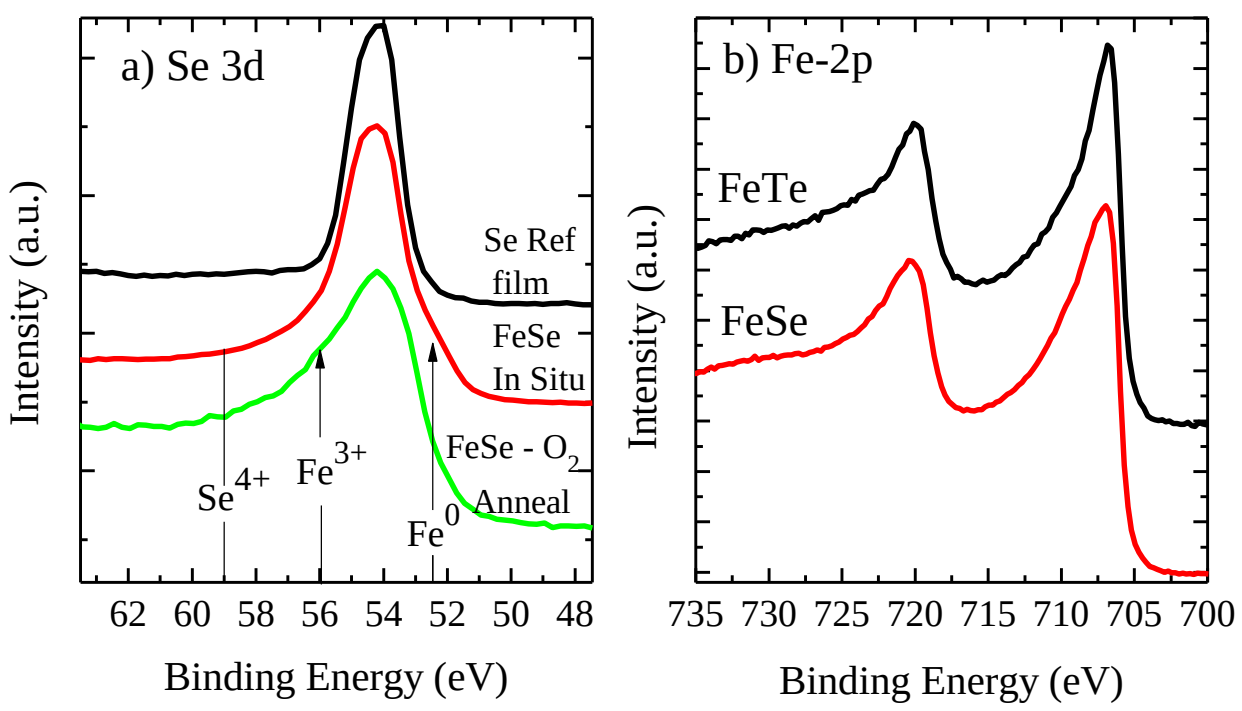

Figure 4: a) Se-3d XPS spectra for an FeSe film grown and measured in-vacuo compared to an $O_2$ annealed FeSe film oxidized. The top spectrum a metallic Se reference film. The arrows represent the binding energies of $Fe^0$ (52.5 eV), $Fe^{3+}$ (56 eV) $Se^{4+}$ (59 eV) from the literature. b) Fe 2p XPS spectra measured in-vacuo following growth of the FeTe and FeSe films.

The Se-3d spectrum for the metallic Se reference film with the binding energy of 54.3 eV is in good agreement with the literature for zero valence Se.[21] The spectrum from the in-vacuo transferred FeSe film also shows a peak maximum located at 54.3 eV, consistent with $Se^0$. The oxygen annealed FeSe film, which turns non-superconducting, exhibits neither a shift of the peak maximum nor appearance of any structure associated with other Se valence states. Specifically, there is no structure located at a binding energy of 59 eV where $Se^{4+}$ would be located.[22] Such absence of the $Se^{4+}$ in the XPS spectra we take as strong evidence that valence changes in the bulk of the film are unlikely to occur. We conclude that regardless of oxygen exposure, Se always remains zero valence throughout the majority of the FeSe films.

The spectra of the FeSe films have a slightly different shape than the spectrum from the Se film. In the spectrum of the in-vacuo FeSe film, a small shoulder at 52.5 eV corresponds to the Fe-3p peak of $Fe^0$.[23] This is consistent with the Fe-2p XPS spectra in Figure 4b), which shows only metallic Fe. The Se 3d spectra of $O_2$ annealed FeSe film exhibits a reduction in intensity at 52.5 eV and increase in intensity at 56 eV which is consistent with our observation of the Fe valence changing from $Fe^0$ to $Fe^{3+}$.[23]

The correlation between observed valence changes and the appearance of superconductivity in FeSe and FeTe films opens some intriguing possibilities. Before discussing them we rule out some trivial explanations. Superconductivity in FeSe films and oxygen-incorporated-FeTe films is a bulk phenomenon, as is evidenced by the Meissner effect measured in both systems[2,9]. The possible impact of oxygen incorporation on the FeTe and FeSe compositions was ascertained by Energy Dispersive X-ray (EDX) measurements. Atomic percentages obtained from EDX (accurate to ± 1%) indicate that the oxygen annealed and air exposed FeTe and FeSe films have no significant deviation from their initial stoichiometry (less than 3% from Fe/X = 1/0.95). These results demonstrate that the onset of superconductivity in FeTe and destruction of superconductivity in FeSe are not the result of sample degradation caused by significant oxygen incorporation. Similarly, the XRD measurements of as-grown and oxygenated FeTe and FeSe films show no change in overall structure and a small change in the lattice constant. Furthermore, the onset and disappearance of superconductivity does not appear to be a specifically film related effect; the films are fully relaxed. Finally, the preliminary DFT calculations on both FeTe and FeSe suggest a common O interstitial site, indicating again similarity of the two systems.[24]

The fundamental observation that incorporated oxygen creates superconductivity in FeTe and destroys it in FeSe opens a few possibilities for identifying the controlling factor for superconductivity. The most notable difference between superconducting FeTeOx and FeSeOx is the chalcogenide valence. However, we know of no particular reason this might control superconductivity and such an underlying cause seems incompatible with the as grown samples and we know of no similar effect in the Fe pnictides. It may be that there is an underlying two dimensional phase diagram with both charge doping and some structural/strain parameter. The magnetic but non-superconducting parent compound (FeTe) would be at the origin, moving an appropriate distance along the charge or strain axes would produce a superconductor ($FeTeO_x$ and FeSe, respectively), as shown in Figure 5. But moving too far in either or a combination of the two would move beyond the superconducting region to a normal metal ($FeSeO_x$). This model seems qualitatively consistent with a wide range of observations in these chalcogenides and the Fe pnictides. However, it is not clear how to make such a picture quantitative at this time.

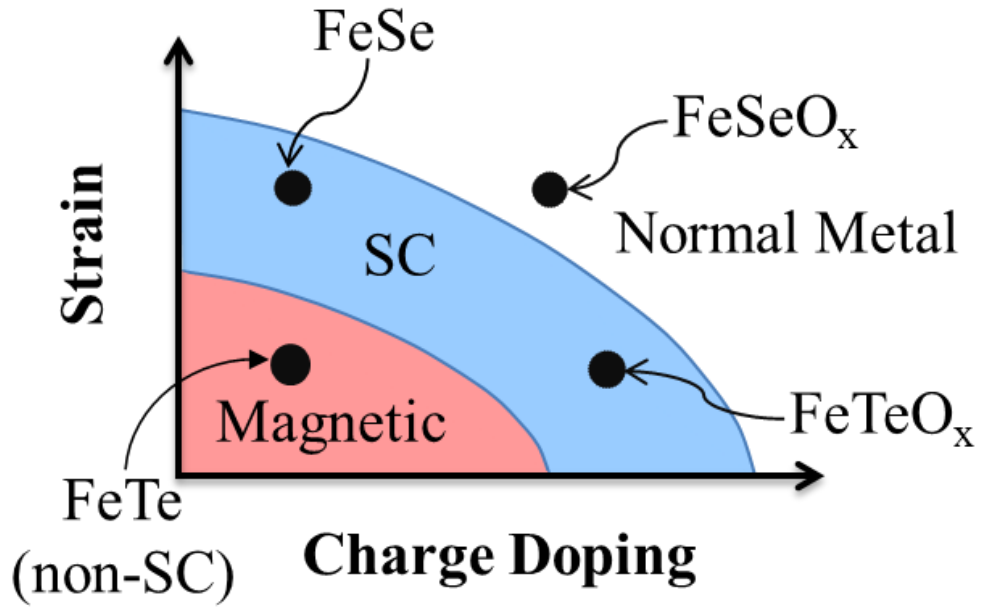

Figure 5: Possible phase diagram for iron chalcogenide superconductors, indicating magnetic, superconducting (SC) and normal metal phases.

Finally, it may simply be that the exact charge state is unimportant for superconductivity and some other feature tips the balance between a magnetic, superconducting, or normal metal ground state. That the exact charge state is not vitally important might be expected for a metal with many bands crossing the Fermi level, but if true this makes it vitally important to identify what parameters are controlling the dominant ground state in each case.

We would like to thank E. Negusse for assistance during absorption measurements at the U4B beam line. This work is supported by the US-DOE through contract # DE-FG02-00ER45801. Use of the National Synchrotron Light Source, Brookhaven National Laboratory was supported by the Office of Science, Office of Basic Energy Sciences, U.S. Department of Energy under Contract No. DEAC02-98CH10886.